\documentclass[12pt,onecolumn,draftcls]{IEEEtran}
\pdfoutput=1

\newif\ifieeetrans
\ieeetransfalse

\newif\ifrevtex
\revtexfalse

\newif\ifendfigure
\endfigurefalse

\ifendfigure
  \usepackage[nomarkers,tablesonly]{endfloat}
\fi

\usepackage{ifpdf}

\usepackage[pdftex]{graphicx}

\usepackage{authblk}
\usepackage{rotating}
\usepackage{tablefootnote}
\usepackage{hyperref}

\begin{document}

\title{Assessing Material Qualities and Efficiency Limits of III-V on Silicon Solar Cells Using External Radiative Efficiency}

\author{Kan-Hua~Lee,
       Kenji~Araki,
  Li Wang, Nobuaki~Kojima, \\
  Yoshio~Ohshita and Masafumi~Yamaguchi}

\ifrevtex
  \affiliation{Toyota Technological Institute, Nagoya, Aichi, Japan, 468-8511}
\else
  \affil{Toyota Technological Institute, Nagoya, Aichi, Japan, 468-8511}
\fi

\ifrevtex
\else
  \maketitle
\fi

\begin{abstract}
The paper presents a quantitative approach to the investigation and comparison of the material qualities of III-V on silicon (III-V/Si) solar cells by using external radiative efficiencies. We use this analysis to predict the limiting efficiencies and evaluate the criteria of material quality in order to achieve high efficiency III-V/Si solar cells. This result yields several implications for the design of high efficiency III-V/Si solar cells.
\end{abstract}

\ifrevtex
  \maketitle
\fi

\ifieeetrans
  \begin{IEEEkeywords}
  Photovoltaic cell
  \end{IEEEkeywords}
\fi

\section{Introduction}
\footnote{This is the peer reviewed version of the article which has been published in final form at \textit{Progress in Photovoltaics: Research and Applications} (\href{http://dx.doi.org/10.1002/pip.2787}{DOI: 10.1002/pip.2787}). This article may be used for non-commercial purposes in accordance with Wiley Terms and Conditions for Self-Archiving.}
\label{sec:intro}
\ifieeetrans
  \IEEEPARstart{I}II-V
\else
  III-V
\fi
multi-junction solar cells currently demonstrate the highest conversion efficiencies among photovoltaic materials by a wide margin\cite{Green:2015bk}. However, high substrate and fabrication costs limit their uses to the applications such as Space or concentrator photovoltaics that are more tolerant to cell cost. Currently the use of III-V multi-junction solar cells using compound semiconductors lattice-matched to GaAs is by far the most mature technology. 
Replacing the germanium or GaAs substrate with silicon is a promising approach to reduce the substrate cost. 
However, III-V materials that are lattice-matched to silicon are very limited, but using other III-V materials that are lattice mismached to silicon is very challenging to maintain their material qualities. Moreover, the qualities of III-V materials that constitute the optimal band-gap combinations for dual-junction (2J) or triple-junction (3J) III-V/Si solar cells are generally less optimized than standard III-V solar cell materials such as InGaP and GaAs\cite{Connolly:2014jm}.
 
Designing and modeling III-V on silicon solar cells has been considered in a number of publications\cite{Connolly:2014jm}\cite{Jain:2014ko}\cite{Jain:2014ca}\cite{Jain:2012jk}. However, these publications mainly focus on the detailed layer design and optimization of III-V/Si solar cells. In this paper, we aim to investigate and review the design of III-V/Si solar cells from a different point of view, with an emphasis on the issues of the material quality of each subcell.

In this work, we choose external radiative efficiency (ERE) as the measure of the material quality of solar cells. A similar approach has been applied to analyze the state-of-the-art solar cells and predict their performances at high concentrations\cite{Green:2011ea}\cite{Chan:2012ej}. We will use this framework to address the following questions of III-V/Si solar cells: What are the essential criteria for III-V/Si solar cells in order to match the efficiencies of state-of-the-art III-V multi-junction cells? How do these criteria compare to the material quality of state-of-the-art III-V/Si solar cells? Would it be acceptable to sacrifice the material quality of III-V cells in exchange for better current-matching in III-V/Si solar cells?

In this paper, we will first describe the modeling approach and assumptions that we use to predict the performance of the solar cells and how we estimate the EREs from reported results for III-V/Si solar cells. We will then present the modeling results of III-V/Si solar cells with our estimated EREs taken into account. Finally we will discuss the implications of these results on designing III-V/Si solar cells.

\section{Modeling Efficiencies of III-V on Silicon Solar Cells}
\label{sec:modeling}
We calculate the I-V characteristics of III-V/Si solar cells based on a detailed balance model\cite{Shockley:1961co,Araujo:1994jk,Nelson:1997fb}.
We first assume the principle of superposition for the total current density of the solar cells, i.e., the total current density is the sum of recombination current density $J_{tot}(V)$ and short-circuit current density $J_{sc}$, which are decoupled from each other:
\begin{equation}
\label{eqn:jv_eq}
J(V)=-J_{sc}+J_{tot}(V)
\end{equation}
$J_{sc}$ can be written as the integration of external quantum efficiencies (EQE) multiplied by the input spectrum over photon energy $E$, namely,
\begin{equation}
\label{eqn:Jsc}
J_{sc}=q \int_0^{\infty} \phi(E) \cdot \mbox{EQE}(E) dE
\end{equation}
where $\phi(E)$ is incident photon spectrum. We assume flat, stepped EQEs in our calculations unless otherwise specified, i.e.,
\begin{equation}
\label{eqn:qe_def}
\mbox{EQE}(E)=
\left\{
\begin{array}{ll}
b , E \geq E_g \\
0 , E < E_g
\end{array}
\right.
\end{equation}
where $E_g$ is the band gap of the material and $b$ is a chosen EQE value. 

As we mentioned in Section~\ref{sec:intro}, external radiative efficiency (ERE) is defined as the fraction of radiative recombination currents against total recombination currents. The total recombination current $J_{tot}(V)$ can thus be related to the radiative recombination current $J_{rad}(V)$ by
\begin{equation}
\label{eqn:eta_r}
J_{tot}(V)=J_{rad}(V)/\eta_r
\end{equation}
where $\eta_r$ is ERE. Note that this equation assumes that the ERE is independent of the level of carrier injection. This may not be valid at the regime of high carrier injection. However, since we focus on III-V/Si solar cells at or near one-sun illumination in this study, assuming a constant ERE is reasonable.

The radiative recombination current density is calculated by a detailed balance approach, which is the total radiative recombination photons escaping from the solar cell per area multiplied by the elementary charge $q$:
\begin{equation}
\label{eqn:gen_planck}
J_{rad}(V)=\frac{2\pi q (n_c^2+n_s^2)}{\mbox{h}^3 \mbox{c}^2}\int_{0}^{\infty} \frac{ a(E) E^2 dE}{\exp\left(\frac{E-qV}{kT}\right)-1}
\end{equation}
where $n_c$ is the refractive index of the solar cell, $n_s$ is the refractive index of the medium over the solar cell, h is Planck's constant, c is the speed of light, k is Boltzmann constant, and $T$ is the absolute temperature. $a(E)$ is the absorptivity of the cell, which is approximated to be equal to $\mbox{EQE}(E)$ in this study.
Details of this model and the derivation of (\ref{eqn:gen_planck}) can be found in \cite{Araujo:1994jk} or \cite{Nelson:1997fb}. (\ref{eqn:gen_planck}) assumes that the all radiative photons can escape from the surface where the medium next to it is a semiconductor. For the surface that exposes to air, only the radiated photons that lie within the light cone $\theta<\sin^{-1}(1/n_c)$ can escape from the surface.
Also, the solar cell is assumed to be infinite and planar with the emission from the edges neglected.

In the calculation of the efficiencies of multi-junction solar cells, the reflections and parasitic transmission losses of the top surface and the interfaces between junctions are neglected.
Also, it is assumed that all absorbed photons can be converted to electrical currents, whereas photons that are not absorbed can fully transmit to the next junction. In other words, different EQE values of a junction are equivalent to different optical thicknesses of a subcell.
With these assumptions we get the the following expression for the photons incident on the $i$-th junction:
\begin{equation}
\label{eqn:phi_i_E}
\phi_i (E)=\phi_{i-1}(E)(1-\mbox{EQE}_{i-1}(E))
\end{equation}
where $\phi_{i-1}(E)$ and $\mbox{EQE}_{i-1}(E)$ are the incident photons and the EQE of the junction stacked above the $i$-th junction. The I-V characteristics of $i$-th subcell can then be calculated by substituting $\phi_{i}(E)$ and $\mbox{EQE}_{i}(E)$ into (\ref{eqn:Jsc}) and (\ref{eqn:gen_planck}). In this work, we only consider two-terminal, series-connected multi-junction solar cells. The I-V of the multi-junction device is thus solved by interpolating the voltages for each current density $J$ for every subcell and adding up the subcell voltages to obtain the I-V characteristics of the multi-junction cell $V_{tot}(J)$, namely,
\begin{equation}
V_{tot}(J)=\sum_{i=1}^N V_i(J)
\end{equation}
where $V_i(J)$ is the voltage of the $i$-th subcell at the current density $J$. The efficiency is defined by the maximum power point of $V_{tot}(J)$ divided by the total power of the illumination spectrum. 
The illuminating spectrum is AM1.5g normalized to 1000 $\mbox{W/cm}^2$ throughout all the calculations in Section~\ref{sec:design}.

Since our main focus in this study is the impact of non-radiative recombinations due to imperfect material qualities, loss mechanisms such as parasitic resistances and optical losses are neglected. Although ERE may depend on the geometry of solar cells\cite{Steiner:2013cc}, the geometry of multi-junction solar cells is fairly standard and therefore can be considered to be a constant factor. 

\begin{figure}[!t]
\centering
\includegraphics[width=2.5in]{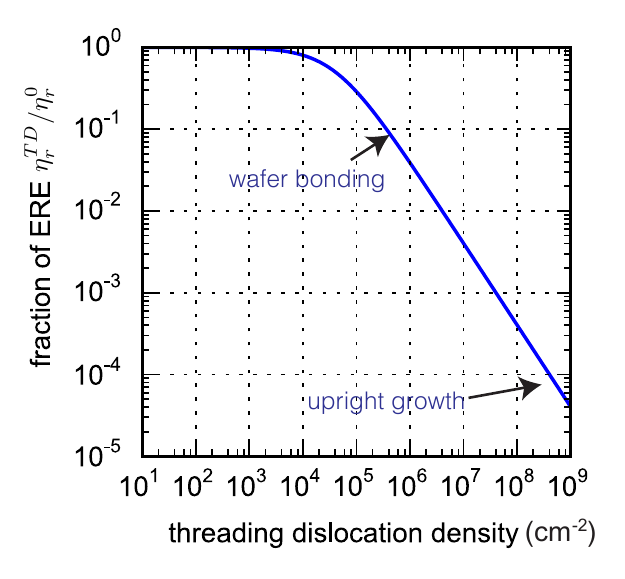}
\caption{Fraction of ERE without threading dislocation ${\eta^{TD}_{r}}/{\eta^{0}_r}$ against threading dislocation densities. The arrows mark the range of ${\eta^{TD}_{r}}/{\eta^{0}_r}$ of III-V subcells on silicon with different fabrication listed in \tablename{~\ref{table:III-V_Si_ERE}}.}
\label{fig:rad_DD_plot}
\end{figure}

Due to the lattice mismatch between most of the III-V compounds and silicon, threading dislocation poses the main challenge to achieve high efficiency III-V/Si solar cells. 
ERE can be related to threading dislocation densities by using a empirical model proposed in \cite{Yamaguchi:1989de}.
First, the reduction of minority carrier lifetime due to threading dislocations can be described by the following equation\cite{Yamaguchi:1989de}:
\begin{equation}
\frac{1}{\tau_{eff}}=\frac{1}{\tau_{rad}}+\frac{1}{\tau_{nr}}+\frac{\pi^3 N_d}{4}
\end{equation}
where $\tau_{eff}$ is the effective minority carrier lifetime, $\tau_{rad}$ is the radiative lifetime, $\tau_{nr}$ is the non-radiative lifetime, and $N_d$ is the threading dislocation density.
After that, by assuming that the carrier injection density is low and the geometry factors are identical, the reduction of ERE due to threading dislocations can be written as \cite{Yamaguchi:1989de}
\begin{equation}
\frac{\eta^{TD}_{r}}{\eta^{0}_r}=\frac{\tau^{TD}_{eff}}{\tau^{0}_{eff}}
\end{equation}
where $\eta^{TD}_{r}$ is the ERE with threading dislocations, $\eta^{0}_{r}$  is the ERE without threading dislocations,  $\tau^{TD}_{eff}$ is the effective minority carrier lifetime with threading dislocations, and $\tau^{0}_{eff}$  is the effective minority carrier lifetime without threading dislocations.
If we only take the degradation of the p-type base layer into account, and assume $\tau^{0}_{eff}=20$ ns \cite{Yamaguchi:1989de}\cite{Andre:2004jy} and $D=80~\mbox{cm}^2 \mbox{/s}$ \cite{Jain:2012jk},  ERE against the threading dislocation density can then be calculated and plotted. See \figurename~\ref{fig:rad_DD_plot}. 
We will discuss this result further in Section~\ref{sec:design}.

\section{EREs of State-of-The Art III/V on Silicon Solar Cells}
\begin{figure}[!t]
\centering
\includegraphics[width=2.5in]{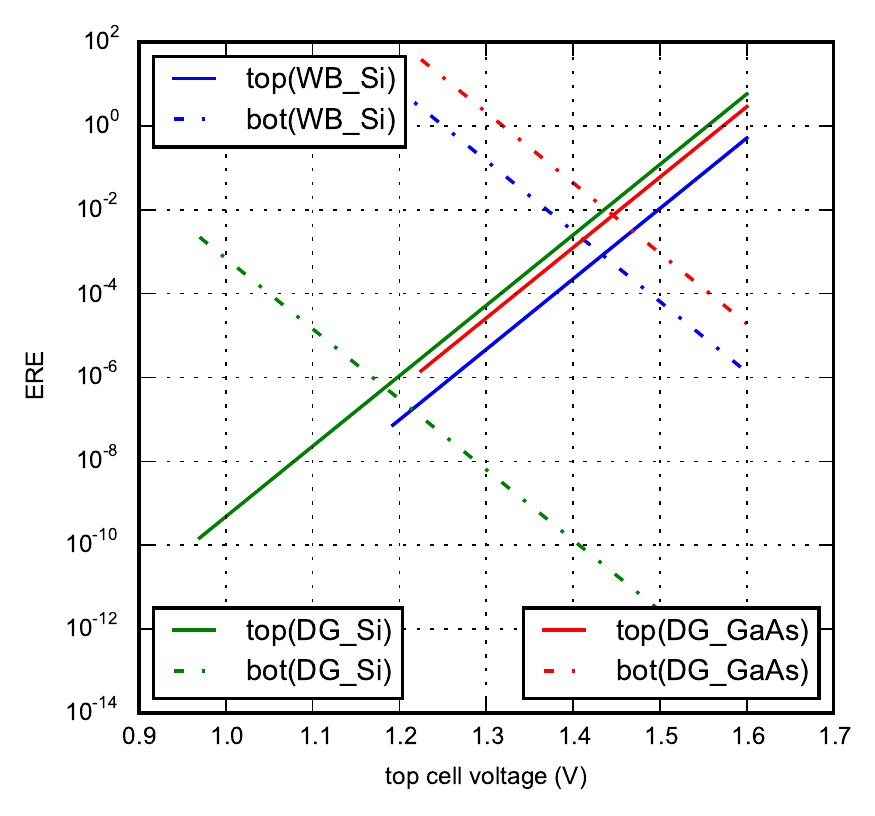}
\caption{Estimated ERE range of 2J InGaP/GaAs on silicon and GaAs substrates reported in \cite{Dimroth:2014jn}. WB\_Si stands for wafer bonding on silicon substrate, DG\_GaAs stands for direct growth on GaAs substrate, and DG\_Si stands for direct growth on silicon.}
\label{fig:dimroth_paper_radeta}
\end{figure}

Based on the theoretical framework described Section~\ref{sec:modeling}, we can analyze the EREs of III-V and III-V/Si solar cells reported in the publications that are relevant to this study.
We use the assumptions underlying the principle of superposition in (\ref{eqn:jv_eq}) to approximate ERE, that is, the value of $J_{sc}$ is
equal to the total recombination current $J_{tot}$ at $V_{oc}$
\begin{equation}
J_{tot}(V_{oc})=J_{sc}
\end{equation}
Following from the definition of ERE in (\ref{eqn:eta_r}), ERE can be written as 
\begin{equation}
\label{eqn:extract_ERE}
\eta_r=\frac{J_{rad}(V_{oc})}{J_{sc}},
\end{equation}
where $J_{rad}(V_{oc})$ and $J_{sc}$ can be calculated from (\ref{eqn:gen_planck}) and (\ref{eqn:Jsc}), respectively. In this way, it only requires EQE, $V_{oc}$, and the illumination spectrum in order to estimate ERE.
The EREs of some state-of-the-art solar cells have been presented in \cite{Green:2011ea} and \cite{Geisz:2013hi}. These are listed in \tablename~\ref{table:stta_ERE}. 
We also calculated the EREs of III-V/Si and III-V solar cells in \cite{Virshup:1985hw}\cite{Amano:1987es} and \cite{Soga:1995bv} using this approach. The results are listed in \tablename~\ref{table:III-V_Si_ERE}.
For single-junction devices, estimating their EREs is straightforward by using (\ref{eqn:extract_ERE}) with measured open circuit voltages and EQEs. 
However, for multi-junction solar cells, we often can only know the open-circuit voltage of the entire device as opposed to those of the individual subcells. We thus need to make a few more assumptions in order to estimate the EREs of the subcells. First, we assume that the $V_{oc}$ of the whole device is the sum of $V_{oc}$ of each subcell, i.e.,
\begin{equation}
\label{eqn:voc_sum}
V_{oc}=\sum_{i=1}^{N}V_{oc}^i
\end{equation}
where $V_{oc}^i$ is the open-circuit voltage of the $i$-th junction of the $N$-junction solar cell. This assumption is reasonable because the current mismatch of the multi-junction cells that we selected to analyze is less than 10\%. After that, we select a reasonable voltage range of top cell's $V_{oc}$ and calculate its corresponding EREs. 
For 2J cells, we can then calculate the bottom cell's $V_{oc}$ and its ERE for each top cell's $V_{oc}$ based on the assumption of (\ref{eqn:voc_sum}).
We use this method to estimate and compare the EREs of three III-V 2J solar cell reported in \cite{Dimroth:2014jn}. These were fabricated by different methods, including one InGaP/GaAs 2J cell on GaAs substrate,  one InGaP/GaAs 2J cell on silicon substrate using wafer bonding and one InGaP/GaAs 2J on silicon substrate using direct growth. The estimated EREs of these 2J cells presented in \cite{Dimroth:2014jn}  are plotted in \figurename~\ref{fig:dimroth_paper_radeta}.

Since it is more likely that the InGaP cell and the GaAs cell have similar EREs, the actual range of EREs can be limited to the region near the intersection of the top and bottom cell's EREs. This is also due to that the y-axis in  \figurename~\ref{fig:dimroth_paper_radeta} is presented in the log scale, and only the regions near the intersections cover the non-negligible range of EREs for both subcells. 
For 2J cells grown on GaAs, the range of the EREs can be further reduced to the left-hand side of the intersection because the ERE of GaAs is generally higher than InGaP. 
From \figurename~\ref{fig:dimroth_paper_radeta}, we can see that the intersection of ERE lines of the 2J cell grown on a GaAs substrate is around $10^{-2}$, which is close to the values of the state-of-the-art GaAs single junction cell listed in \tablename~\ref{table:stta_ERE}. The ERE of the wafer-bonded 2J cell drops to around $10^{-3}$, whereas the ERE of the 2J cell epitaxially grown on a silicon substrate is reduced to only around $10^{-6}$. These estimated EREs are also labeled in \figurename~\ref{fig:rad_DD_plot}. The corresponding dislocation density of direct-growth 2J on silicon cells of these EREs match the value reported in \cite{Dimroth:2014jn}, which is around $10^{8}~\mbox{cm}^{-2}$.

\begin{figure}[!t]
\centering
\includegraphics[width=2.5in]{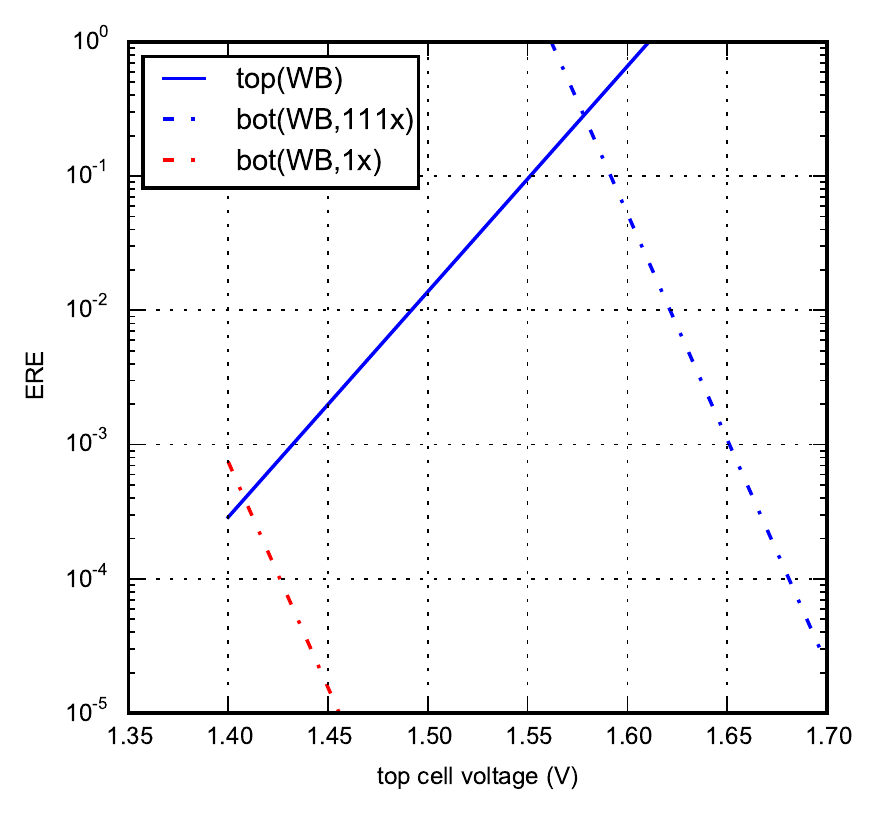}
\caption{Estimated ERE range of 3J InGaP/GaAs/Si solar cells fabricated by wafer bonding as reported in \cite{Essig:2015dw}. Both the estimated EREs of the cell tested at one sun (1x) and 111 suns (111x) are plotted.}
\label{fig:essig_paper_radeta}
\end{figure}

This ERE estimation method was also applied to the case of GaInP/GaAs/Si 3J cells\cite{Essig:2015dw}. Since we have three cells with unknown EREs, we have to make an additional assumption that the top and middle cells have the same ERE so that we can make a two-dimensional plot as in \figurename~\ref{fig:dimroth_paper_radeta}.
\figurename~\ref{fig:essig_paper_radeta} shows the estimated EREs of the 3J wafer-bonded InGaP/GaAs/Si solar cell measured at one sun and 112 suns. From the results in \figurename~\ref{fig:dimroth_paper_radeta}, we know that the EREs of InGaP/GaAs 2J solar cell are around $10^{-3}$. Because the cells in \figurename~\ref{fig:dimroth_paper_radeta} and \figurename~\ref{fig:essig_paper_radeta} came from the same research group, we may assume that the III-V subcells in \figurename~\ref{fig:essig_paper_radeta} have similar EREs at one sun. Thus the ERE of the silicon bottom cell is then around $10^{-4}$. When the cell is illuminated with concentrated sunlight, the EREs of the subcells can be raised by several orders of magnitude. This may be due to the saturation of defect states or the reduction of etendue loss\cite{Hirst:2010ch}.
We also estimated the EREs of the III-V/Si solar cells presented in \cite{Yang:2014kz}. Although the EQE of the subcells are not reported in \cite{Yang:2014kz}, we infer that the EREs of the subcells are similar to the results in \figurename~\ref{fig:essig_paper_radeta} according to the reported values of short circuit current and open-circuit voltages. These results are all listed in \tablename~\ref{table:III-V_Si_ERE}.

\section{Design Considerations of III-V/Si Solar Cells} \label{sec:design}
\begin{figure}[!t]
\centering
\includegraphics[width=2.5in]{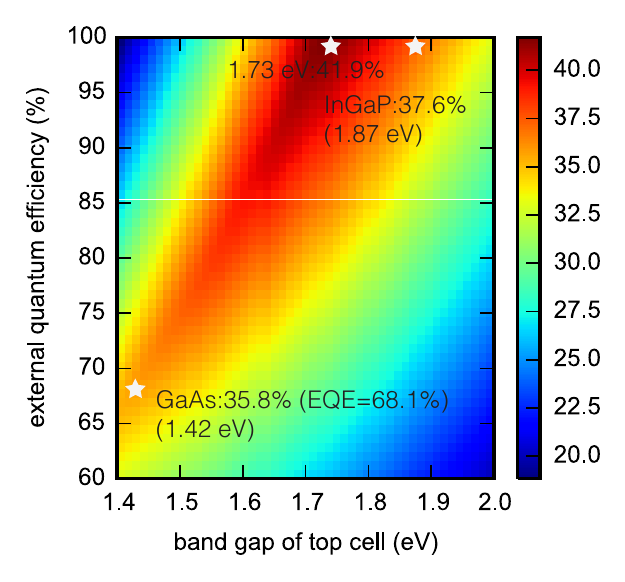}
\caption{Efficiency contours of 2J solar cells with silicon bottom cell as a function of band gap and EQE of the top cell. The EREs are set to be 1 for both subcells. The EQE of the silicon bottom cell is 100\%. The color bar is efficiency (\%).}
\label{fig:rad_limit_2J}
\end{figure}

This section presents the estimated efficiencies of multi-junction solar cells with different EREs and their implications for designing III-V/Si solar cells. Starting from the radiative-limit case, i.e. ERE is 1 and no optical losses for all subcells, \figurename~\ref{fig:rad_limit_2J} shows the efficiency contour of 2J III-V/Si solar cells as a function of the band gap and the EQE of the top cell. Altering EQE in this calculation is equivalent to different optical thicknesses of the top cell, as we mentioned in (\ref{eqn:phi_i_E}). In the calculations throughout this section, the band gap of silicon is assumed to be 1.12 eV and the EQE is 100\%. The result shows that the maximum predicted efficiency, with 1.73 eV as the top cell band gap, is 41.9\%. The optimal band gap of the top cell can potentially be achieved by using AlGaAs, Ga(As)PN or AlInGaP, but it remains challenging to achieve high quality in these materials.

From the point of view of material quality, GaAs is a favored option for the top junction, but its band gap is too close to silicon and thus makes silicon as the current-limiting junction. 
Reducing the optical thickness of GaAs can mitigate the current-mismatch and raise the efficiency. 
As shown in \figurename~\ref{fig:rad_limit_2J}, the optimal EQE of the GaAs top cell is around 68.1\%, which gives 35.8\% efficiency of the 2J device.
Another option is choosing InGaP (1.87eV) as the top junction on silicon cell. The limiting efficiency of this configuration is 37.6\%, which is even higher than the limiting efficiency of GaAs/Si. With this configuration, the current-limiting junction then becomes the top cell, providing the opportunity to use thinner silicon junction to reduce the recombination current.

\begin{figure}[!t]
\centering
\includegraphics[width=2.5in]{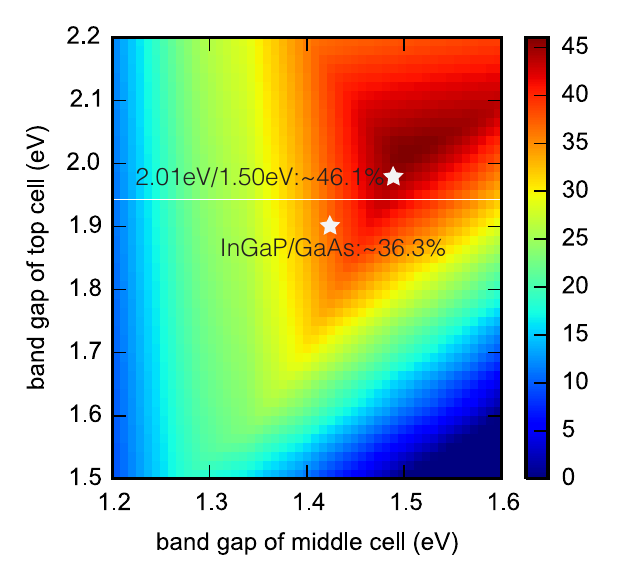}
\caption{Efficiency contours of 3J solar cell with silicon bottom cell as a function of top and middle cell band gaps. The EREs and EQEs of all subcells are assumed to be 100\% in this calculation. The color bar is efficiency (\%).}
\label{fig:rad_limit_3J}
\end{figure}

\begin{figure}[!t]
\centering
\includegraphics[width=2.5in]{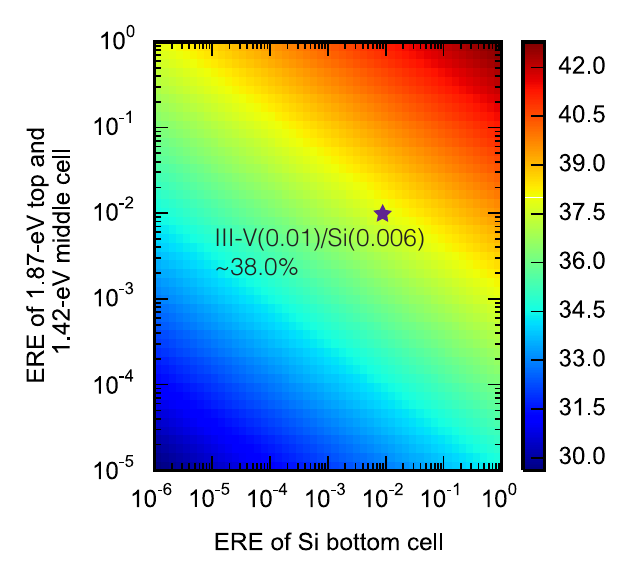}
\caption{Efficiency contours of 3J InGaP/GaAs/Si solar cell as a function of the EREs of III-V and silicon bottom cell junctions. The EREs of top and middle junctions are assumed to be identical. The EQE of the bottom cell is 100\%, and the EQEs of the top and middle cells are 82.6\%, which are optimal EQE values for this configuration. The color bar is efficiency (\%). The EREs on x- and y-axis are in numerics.}
\label{fig:3J_si_vary_radeta_conv}
\end{figure}

\begin{figure}[!t]
\centering
\includegraphics[width=3in]{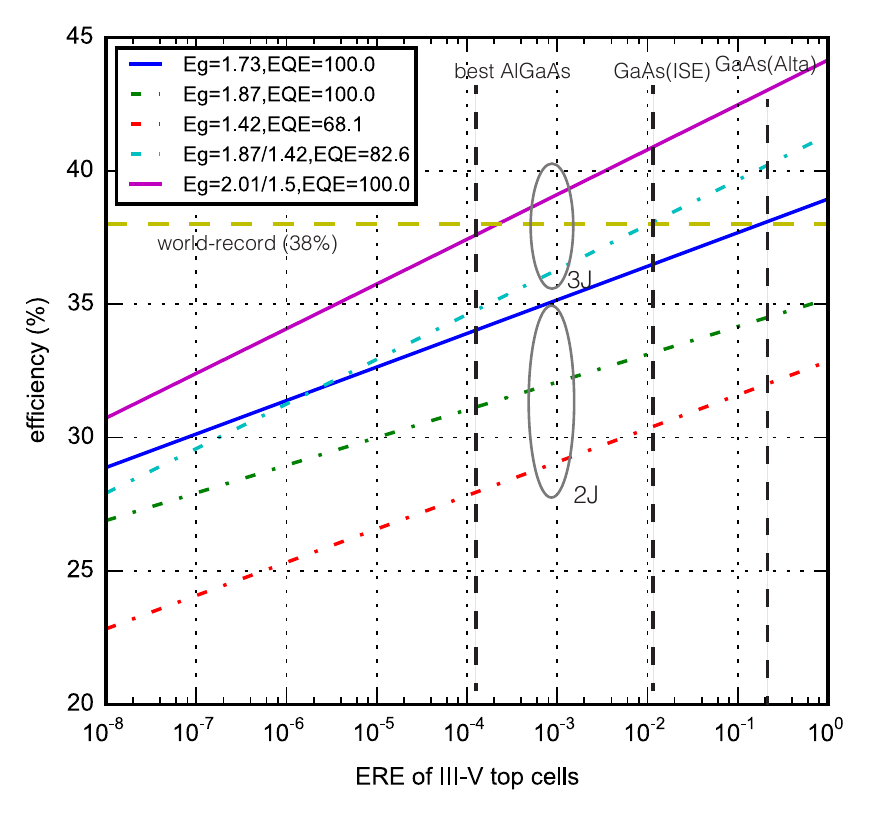}
\caption{
Predicted efficiency limits of several band-gap configurations of III-V subcell on silicon bottom cell against the EREs of III-V subcells. The optimal band-gap combinations are plotted in solid lines and sub-optimal combinations are plotted in broken lines.
The EQEs in these calculation are chosen to give the maximum overall efficiencies.
The band-gap configurations (eV) and the EQEs(\%) of the top cells are described in the legend.
Also, the EQE of the silicon bottom cell is assumed to be 100\% and the ERE is assumed to be 0.006, which is the state-of-the-art value reported in \cite{Green:2011ea}. The EREs on the x-axis are presented in numerics.}
\label{fig:rad_limit_3J_profile}
\end{figure}

\figurename~\ref{fig:rad_limit_3J} shows the efficiency contours of 3J III-V/Si solar cells against the top and middle cell's band gaps. All of the subcells are assumed to have 100\% EQEs. The optimal band-gap combination for the top two junctions is 2.01 eV and 1.50 eV, which gives a limiting efficiency of 46.1\%. Ternary or quaternary compounds such as AlGaAs/AlGaAs, InGaP/GaAsP, (Al)InGaP/InGa(As)P , (Al)InGaP/AlGaAs, and GaPN/GaAsPN are candidates for this optimal band-gap configuration.
Using conventional InGaP/GaAs on silicon can only achieve 36.3\% efficiency at this radiative limit because of current mismatch between the InGaP/GaAs top cell and the silicon bottom cell. As in the case of GaAs on silicon cells, reducing the optical thicknesses of InGaP/GaAs could yield better current-matching and therefore higher efficiency. Our calculations show that the optimal EQEs for InGaP and GaAs subcells are around 82.6\%, which gives a limiting efficiency of 43.3\%.
This optimal EQE value will be used in the subsequent calculations for InGaP/GaAs/Si solar cells, the results of which are shown in \figurename~\ref{fig:3J_si_vary_radeta_conv} and \figurename~\ref{fig:rad_limit_3J_profile}.

\begin{figure}[!t]
\centering
\includegraphics[width=2.5in]{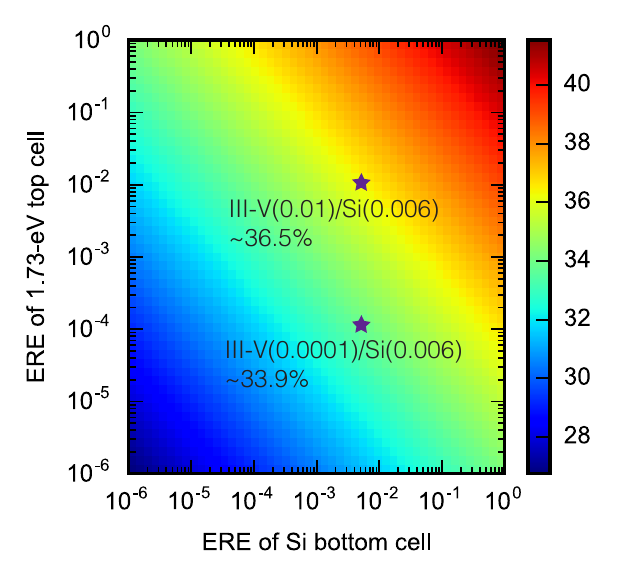}
\caption{Efficiency contours of 2J solar cells as a function of the EREs of the 1.73-eV top and the silicon bottom cell. The EQEs of all subcells are assumed to be 100\% in this calculation. The stars point to the ERE values of state-of-the-art III-V and silicon materials.  The color bar is efficiency (\%). The EREs on x- and y-axis are in numerics.}
\label{fig:1.7eVtop_si_vary_radeta}
\end{figure}

Next we consider how the EREs affect the efficiencies and designs of the solar cells. \figurename~\ref{fig:1.7eVtop_si_vary_radeta} shows the efficiency contours for a 1.73-eV top cell on a silicon bottom cell as a function of the top cell and bottom cell's EREs. Both of the subcells are assumed to have 100\% EQE. 
Based on \tablename~\ref{table:stta_ERE}, the EREs of the state-of-the-art silicon solar cell are around 0.006. For GaAs solar cells, although the cell made by Alta Devices can achieve an ERE of 0.225, this device adopts different cell geometries from the conventional solar cells\cite{Miller:2012fd}. We therefore select the ERE value of ISE's GaAs solar cell as the state-of-the-art value. With these ERE values of each subcell, the limiting efficiency of this 2J cell is 36.5\%. 
Because achieving the quality of state-of-the-art GaAs is still challenging for the candidate 1.73-eV III-V materials, a more realistic estimate is considering a top cell that matches the EREs of the AlGaAs reported in \cite{Virshup:1985hw} or \cite{Amano:1987es}, which are both around $10^{-4}$. The limiting efficiency of this 2J cell is then 33.9\%

A similar calculation was performed for the case of 3J cells.
 \figurename~\ref{fig:3J_si_vary_radeta_conv} is a plot of the efficiency contours of 3J InGaP/GaAs/Si solar cells against the ERE of the III-V junction and the silicon junction.
This calculation assumes that the EREs of the top and the middle cells are identical.
This is a realistic assumption considering the recent improvements in the EREs of InGaP cells\cite{Geisz:2013hi}\cite{Geisz:2014ht}. 
Note that the EQEs of the top and middle cells are chosen to be 82.6\%, which are the optimal values for the conversion efficiencies.
This result shows that the efficiency of InGaP/GaAs/Si could be close to current one-sun world record 3J cell (37.9 \%) \cite{Green:2015bk} if the material quality of every subcell can match the state-of-the-art performances.
However, since this calculation ignores other loss mechanisms such as optical loss or resistance loss, this means that the EREs of these subcells have to be further improved in order to match the performance of current one-sun world-record 3J cell, even without the presence of threading dislocations caused by the lattice mismatch.  

\begin{figure}[!t]
\centering
\includegraphics[width=2.5in]{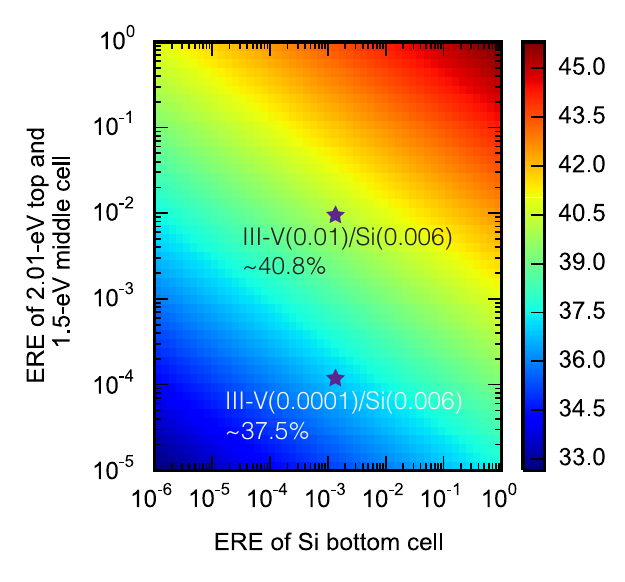}
\caption{Efficiency contours of 3J 2.01eV/1.50eV/Si solar cell as a function of the EREs of III-V and silicon bottom cell junctions. The EREs of top and middle junctions are assumed to be identical. The EQEs of all the subcells are set to be 100\%. The color bar is efficiency in percent (\%). The EREs on x- and y-axis are presented in numerics.}
\label{fig:3J_si_vary_radeta_optimal}
\end{figure}

\figurename~\ref{fig:3J_si_vary_radeta_optimal} shows the results of a similar calculation but with optimal band gaps for the top two junctions. This calculation assumes that all subcells have 100\% EQEs. 
This result shows that the efficiency of a 3J cell can achieve 40.8\% if the III-V materials that constitute this band-gap configuration can match the quality of state-of-the-art GaAs. 
Since the EREs of these candidate materials are still far less than GaAs, a more practical efficiency prospect may be estimated using the EREs of the AlGaAs cells listed in \tablename~\ref{table:III-V_Si_ERE}, which is around $10^{-4}$. This gives an efficiency of around 37.5\%.

As we mentioned earlier, one dilemma in designing III-V/Si solar cells is that the materials that give better current-matching to silicon have poorer material quality, whereas materials with better qualities do not give perfect current-matching. 
By using ERE to quantify the material quality, this issue can be addressed in a more systematic way. \figurename~\ref{fig:rad_limit_3J_profile} shows calculated efficiencies against EREs of several different band-gap configurations of III-V/Si solar cells.
In this calculation, the ERE of the silicon bottom cell is assumed to match the state-of-the-art value (0.006), as listed in \tablename~\ref{table:stta_ERE}. 
The EQE of the silicon bottom cell are set to be 100\%. In the cases of 3J cells, the EREs of the top two junctions are assumed to be identical. Also, we select the EQEs of the III-V junctions that give the best conversion efficiency. These EQE values are listed in the legend of \figurename~\ref{fig:rad_limit_3J_profile}. Note that this is equivalent to optimizing the optical thicknesses of the III-V top cells.
By comparing the efficiency profiles of optimal band-gap combinations and conventional InGaP/GaAs, we see that optimal band-gap combinations can improve efficiencies as long as the ratio of these two EREs is within near $10^{-2}$. For example, as shown in \tablename~\ref{table:stta_ERE}, the ERE of the state-of-the-art conventional GaAs solar cell is around $10^{-2}$. Therefore, in order to match the efficiencies of InGaP/GaAs/Si with 0.01-ERE III-V subcells, and give an optimal band-gap combinations, the EREs of the III-V materials should be close to $2\times10^{-4}$.
According to the EREs achieved by some AlGaAs solar cells reported in \cite{Virshup:1985hw}, \cite{Amano:1987es}, achieving this ERE value may be a realistic target. 

As mentioned before, this calculation assumes no optical loss in the subcells. In \tablename~\ref{table:stta_ERE} and  \ref{table:III-V_Si_ERE}, we list $\eta_{opt}$, which is defined as the ratio between the measured $J_{sc}$ of the solar cell and its ideal $J_{sc}$.  We can see that $\eta_{opt}$ of state-of-the-art silicon and GaAs solar cell can achieve more than 90\%, and the $\eta_{opt}$ of the InGaP is around 82\%. For AlGaAs, the best $\eta_{opt}$ value is around ~81\%, which is close to the $\eta_{opt}$ of the InGaP.
Therefore, as an approximation, one can simply multiply the y-axis of \figurename~\ref{fig:rad_limit_3J_profile} by 80\% to take into account the parasitic optical losses of state-of-the-art cells. In this way, the ERE criteria for AlGaAs/AlGaAs/Si to outperform InGaP/GaAs/Si with optical loss considered would be close to the ERE criteria without parasitic optical loss. However, if the optimal-band-gap materials have much less $\eta_{opt}$ than InGaP and GaAs, we expect that this tolerance of the EREs for optimal band-gap materials would reduce.

\section{Conclusion}
By using a detailed balance model and EREs, we reviewed and compared the material qualities of several different single- or multi-junction III-V/Si solar cells. We also estimated the efficiencies of III-V/Si solar cells with various band-gap configurations and EREs. For InGaP/GaAs/Si solar cells, our calculation shows that, while all of the subcells can match the state-of-the-art EREs, they still cannot match the efficiency of the current one-sun 3J world record. Achieving this is more likely with optimal band gaps of the two junctions, but improving the material qualities of the candidate III-V compounds will be challenging.
We also made relative comparison between InGaP/GaAs/Si and the optimal band-gap configuration, 2.01eV/1.50eV/Si. Our calculation indicates that choosing III-V materials with optimal band-gap combinations with silicon can yield better efficiency compared to InGaP/GaAs, as long as the EREs of these III-V materials are within two order of magnitude less than the EREs of InGaP/GaAs. The estimated EREs of previously reported AlGaAs solar cell suggest that this criteria may be achievable.
\begin{table}[!t]
\renewcommand{\arraystretch}{1.3}
\caption{External radiative efficiencies of state-of-the-art silicon and GaAs solar cells. The data in the first four rows are excerpted from \cite{Green:2011ea}, whereas the last two rows are excerpted from \cite{Geisz:2013hi}. $\eta_{opt}$ is the ratio between the measured $J_{sc}$ of a cell and its ideal $J_{sc}$ calculated by using (\ref{eqn:qe_def}) and assuming 100\% EQE. }
\label{table:stta_ERE}
\centering
\footnotesize
\begin{tabular}{cccccc}
\hline
Device & $V_{oc} (\mbox{mV})$ & $J_{sc} (\mbox{mA}/\mbox{cm}^2)$ & $\eta_{opt} (\%)$  & $\eta$ (\%) & ERE \\
\hline
Si UNSW\tablefootnote{University of New South Wales}& 706 & 42.7 & 97.5 & 25.0 & 0.0057\\
Si SPWR\tablefootnote{SunPower Corporation} & 721 & 40.5 & 92.4  &24.2 & 0.0056 \\
GaAs Alta \tablefootnote{Alta Devices} & 1107 & 29.6  &  92.3   & 27.6 & 0.225 \\
GaAs ISE \tablefootnote{Fraunhofer Institute for Solar Energy Systems} & 1030 & 29.8 &  92.9 &26.4 & 0.0126 \\
InGaP NREL \tablefootnote{National Renewable Energy Laboratory}  (conventional) & 1406 & 14.8 & 79.9 & 18.4 & 0.0032 \\
InGaP NREL (inverted rear-hetero)&  1458 & 16.0 &  82.6 & 20.7 & 0.0871  \\
\hline
\end{tabular}
\end{table}

\begin{table*}[!t]
\renewcommand{\arraystretch}{1.3}
\caption{Extracted EREs from selected publications. In the column \textit{Fabrication}, UG, MBE, and MOCVD stand for upright growth, molecular beam epitaxy, and metal-organic chemical vapor deposition, respectively. $\eta$ is conversion efficiency. $\eta_{opt}$ is the ratio between the measured $J_{sc}$ of a cell and its ideal $J_{sc}$ calculated by using (\ref{eqn:qe_def}) and assuming 100\% EQE. For multi-junction devices, we use the ideal $J_{sc}$ of top cells as the denominator for estimating $\eta_{opt}$.}
\label{table:III-V_Si_ERE}
\centering
\tiny
\begin{tabular}{ccccccccccccc}
\hline
Device & $\eta$(\%)  & Spectrum & $V_{oc} (\mbox{mV})$ & $J_{sc} (\mbox{mA}/\mbox{cm}^2)$ & $\eta_{opt} (\%)$ & Fabrication & Year & ERE  & Reference\\
\hline
AlGaAs(1.64eV) on GaAs (1J) &  19.2 & AM2 & 1.18 & 14.5 & 80.0 & MBE, UG & 1985 & $10^{-4}$  & \cite{Virshup:1985hw} \\
AlGaAs(1.79eV) on GaAs (1J) &  14.6 & AM1.5 & 1.28 & 16.2 & 81.4  & MBE, UG & 1987 & $10^{-4}$  & \cite{Amano:1987es} \\
AlGaAs(1.54eV) on Si (2J) & 20.6 & AM0 & 1.51 & 23.0  & 67.1 & MOCVD, UG & 1996 & AlGaAs: $10^{-7}$ & \cite{Soga:1995bv} \\
                             &          &          &         &           &              &              &           & Si: $10^{-6}$           &             \\
InGaP/GaAs on GaAs (2J) & 27.1 & AM1.5g & 2.45 & 13.15 & 74.2 & MOCVD, UG  & 2014 &  $10^{-2}$ & \cite{Dimroth:2014jn} \\ 
InGaP/GaAs on Si (2J) & 26.0 & AM1.5g & 2.39 & 12.70 & 71.7 & Wafer Bonding  & 2014 & $10^{-3}$ & \cite{Dimroth:2014jn} \\ 
InGaP/GaAs on Si (2J) & 16.4 & AM1.5g & 1.94 & 11.2 & 63.2 & MOCVD, UG  & 2014 &  $10^{-6}$ & \cite{Dimroth:2014jn} \\ 
GaInP/GaAs/Si (3J) & 27.2 & AM1.5d@1x & 2.89 & 11.2 & 73.8 & Wafer Bonding & 2013 & $10^{-3}\sim10^{-4}$ & \cite{Essig:2015dw} \\
GaInP/GaAs/Si (3J) & 30.0 & AM1.5d@111x & 3.4  &1125.4 & 66.8 & Wafer Bonding & 2013 & $\sim10^{-1}$ & \cite{Essig:2015dw} \\ 
GaInP/GaAs/Si (3J) & 27.3 & AM1.5g & III-V:2.23 & III-V:13.7 & 77.3 & Metal Interconnect & 2014 & $10^{-3}\sim10^{-4}$ & \cite{Yang:2014kz} \\ 
                  &      &        & Si:0.49 & Si: 6.88 &       &             &  & $10^{-3}\sim10^{-4}$ & \cite{Yang:2014kz} \\ 
\hline
\end{tabular}
\end{table*}

\section*{Acknowledgement}
The authors would like to thank Japan New Energy and Industrial Technology Development Organization (NEDO) for supporting this research (NEDO 15100731-0).

\bibliographystyle{IEEEtran}
\bibliography{rad_eta_lib}

\begin{thebibliography}{10}
\providecommand{\url}[1]{#1}
\csname url@samestyle\endcsname
\providecommand{\newblock}{\relax}
\providecommand{\bibinfo}[2]{#2}
\providecommand{\BIBentrySTDinterwordspacing}{\spaceskip=0pt\relax}
\providecommand{\BIBentryALTinterwordstretchfactor}{4}
\providecommand{\BIBentryALTinterwordspacing}{\spaceskip=\fontdimen2\font plus
\BIBentryALTinterwordstretchfactor\fontdimen3\font minus
  \fontdimen4\font\relax}
\providecommand{\BIBforeignlanguage}[2]{{%
\expandafter\ifx\csname l@#1\endcsname\relax
\typeout{** WARNING: IEEEtran.bst: No hyphenation pattern has been}%
\typeout{** loaded for the language `#1'. Using the pattern for}%
\typeout{** the default language instead.}%
\else
\language=\csname l@#1\endcsname
\fi
#2}}
\providecommand{\BIBdecl}{\relax}
\BIBdecl

\bibitem{Green:2015bk}
\BIBentryALTinterwordspacing
M.~A. Green, K.~Emery, Y.~Hishikawa, W.~Warta, and E.~D. Dunlop,
  ``\BIBforeignlanguage{English}{{Solar cell efficiency tables (version
  46)}},'' \emph{\BIBforeignlanguage{English}{Progress in Photovoltaics:
  Research and Applications}}, vol.~23, no.~7, pp. 805--812, Jun. 2015.
  [Online]. Available: \url{http://doi.wiley.com/10.1002/pip.2637}
\BIBentrySTDinterwordspacing

\bibitem{Connolly:2014jm}
\BIBentryALTinterwordspacing
J.~P. Connolly, D.~Mencaraglia, C.~Renard, and D.~Bouchier,
  ``\BIBforeignlanguage{English}{{Designing III-V multijunction solar cells on
  silicon}},'' \emph{\BIBforeignlanguage{English}{Progress in Photovoltaics:
  Research and Applications}}, vol.~22, no.~7, pp. 810--820, Jan. 2014.
  [Online]. Available: \url{http://doi.wiley.com/10.1002/pip.2463}
\BIBentrySTDinterwordspacing

\bibitem{Jain:2014ko}
\BIBentryALTinterwordspacing
N.~Jain and M.~K. Hudait, ``{Design and Modeling of Metamorphic Dual-Junction
  InGaP/GaAs Solar Cells on Si Substrate for Concentrated Photovoltaic
  Application},'' \emph{IEEE Journal of Photovoltaics}, vol.~4, no.~6, pp.
  1683--1689, Oct. 2014. [Online]. Available:
  \url{http://ieeexplore.ieee.org/lpdocs/epic03/wrapper.htm?arnumber=6891139}
\BIBentrySTDinterwordspacing

\bibitem{Jain:2014ca}
\BIBentryALTinterwordspacing
------, ``{III{\textendash}V Multijunction Solar Cell Integration with Silicon:
  Present Status, Challenges and Future Outlook},'' \emph{Energy Harvesting and
  Systems}, vol.~1, no. 3-4, pp. 121--145, 2014. [Online]. Available:
  \url{http://www.degruyter.com/view/j/ehs.2014.1.issue-3-4/ehs-2014-0012/ehs-2014-0012.xml}
\BIBentrySTDinterwordspacing

\bibitem{Jain:2012jk}
\BIBentryALTinterwordspacing
------, ``{Impact of Threading Dislocations on the Design of GaAs and
  InGaP/GaAs Solar Cells on Si Using Finite Element Analysis},'' \emph{IEEE
  Journal of Photovoltaics}, vol.~3, no.~1, pp. 528--534, Dec. 2012. [Online].
  Available:
  \url{http://ieeexplore.ieee.org/lpdocs/epic03/wrapper.htm?arnumber=6294426}
\BIBentrySTDinterwordspacing

\bibitem{Green:2011ea}
\BIBentryALTinterwordspacing
M.~A. Green, ``\BIBforeignlanguage{English}{{Radiative efficiency of
  state-of-the-art photovoltaic cells}},''
  \emph{\BIBforeignlanguage{English}{Progress in Photovoltaics: Research and
  Applications}}, vol.~20, no.~4, pp. 472--476, Sep. 2011. [Online]. Available:
  \url{http://doi.wiley.com/10.1002/pip.1147}
\BIBentrySTDinterwordspacing

\bibitem{Chan:2012ej}
\BIBentryALTinterwordspacing
N.~L.~A. Chan, N.~J. Ekins-Daukes, J.~G.~J. Adams, M.~P. Lumb, M.~Gonzalez,
  P.~P. Jenkins, I.~Vurgaftman, J.~R. Meyer, and R.~J. Walters, ``{Optimal
  Bandgap Combinations--Does Material Quality Matter?}'' \emph{IEEE Journal of
  Photovoltaics}, vol.~2, no.~2, pp. 202--208, Mar. 2012. [Online]. Available:
  \url{http://ieeexplore.ieee.org/lpdocs/epic03/wrapper.htm?arnumber=6133321}
\BIBentrySTDinterwordspacing

\bibitem{Shockley:1961co}
\BIBentryALTinterwordspacing
W.~Shockley and H.~J. Queisser, ``\BIBforeignlanguage{English}{{Detailed
  Balance Limit of Efficiency of p-n Junction Solar Cells}},''
  \emph{\BIBforeignlanguage{English}{Journal of Applied Physics}}, vol.~32,
  no.~3, pp. 510--519, 1961. [Online]. Available:
  \url{http://scitation.aip.org/content/aip/journal/jap/32/3/10.1063/1.1736034}
\BIBentrySTDinterwordspacing

\bibitem{Araujo:1994jk}
\BIBentryALTinterwordspacing
G.~L. Ara{\'u}jo and A.~Mart{\'\i}, ``\BIBforeignlanguage{English}{{Absolute
  limiting efficiencies for photovoltaic energy conversion}},''
  \emph{\BIBforeignlanguage{English}{Solar Energy Materials and Solar Cells}},
  vol.~33, no.~2, pp. 213--240, 1994. [Online]. Available:
  \url{http://linkinghub.elsevier.com/retrieve/pii/0927024894902097}
\BIBentrySTDinterwordspacing

\bibitem{Nelson:1997fb}
\BIBentryALTinterwordspacing
J.~Nelson, J.~Barnes, N.~Ekins-Daukes, B.~Kluftinger, E.~Tsui, K.~Barnham,
  C.~T. Foxon, T.~Cheng, and J.~S. Roberts,
  ``\BIBforeignlanguage{English}{{Observation of suppressed radiative
  recombination in single quantum well p-i-n photodiodes}},''
  \emph{\BIBforeignlanguage{English}{Journal of Applied Physics}}, vol.~82,
  no.~12, pp. 6240--6246, 1997. [Online]. Available:
  \url{http://scitation.aip.org/content/aip/journal/jap/82/12/10.1063/1.366510}
\BIBentrySTDinterwordspacing

\bibitem{Steiner:2013cc}
\BIBentryALTinterwordspacing
M.~A. Steiner, J.~F. Geisz, I.~Garc{\'\i}a, D.~J. Friedman, A.~Duda, and S.~R.
  Kurtz, ``\BIBforeignlanguage{English}{{Optical enhancement of the
  open-circuit voltage in high quality GaAs solar cells}},''
  \emph{\BIBforeignlanguage{English}{Journal of Applied Physics}}, vol. 113,
  no.~12, p. 123109, 2013. [Online]. Available:
  \url{http://scitation.aip.org/content/aip/journal/jap/113/12/10.1063/1.4798267}
\BIBentrySTDinterwordspacing

\bibitem{Yamaguchi:1989de}
\BIBentryALTinterwordspacing
M.~Yamaguchi, C.~Amano, and Y.~Itoh, ``{Numerical analysis for high-efficiency
  GaAs solar cells fabricated on Si substrates},'' \emph{Journal of Applied
  Physics}, pp. 915--919, 1989. [Online]. Available:
  \url{http://scitation.aip.org/content/aip/journal/jap/66/2/10.1063/1.343520}
\BIBentrySTDinterwordspacing

\bibitem{Andre:2004jy}
\BIBentryALTinterwordspacing
C.~L. Andre, J.~J. Boeckl, D.~M. Wilt, A.~J. Pitera, M.~L. Lee, E.~A.
  Fitzgerald, B.~M. Keyes, and S.~A. Ringel,
  ``\BIBforeignlanguage{English}{{Impact of dislocations on minority carrier
  electron and hole lifetimes in GaAs grown on metamorphic SiGe substrates}},''
  \emph{\BIBforeignlanguage{English}{Applied Physics Letters}}, vol.~84,
  no.~18, pp. 3447--3449, 2004. [Online]. Available:
  \url{http://scitation.aip.org/content/aip/journal/apl/84/18/10.1063/1.1736318}
\BIBentrySTDinterwordspacing

\bibitem{Geisz:2013hi}
\BIBentryALTinterwordspacing
J.~F. Geisz, M.~A. Steiner, I.~Garc{\'\i}a, S.~R. Kurtz, and D.~J. Friedman,
  ``\BIBforeignlanguage{English}{{Enhanced external radiative efficiency for
  20.8\% efficient single-junction GaInP solar cells}},''
  \emph{\BIBforeignlanguage{English}{Applied Physics Letters}}, vol. 103,
  no.~4, pp. 041\,118--6, 2013. [Online]. Available:
  \url{http://scitation.aip.org/content/aip/journal/apl/103/4/10.1063/1.4816837}
\BIBentrySTDinterwordspacing

\bibitem{Virshup:1985hw}
\BIBentryALTinterwordspacing
G.~F. Virshup, C.~W. Ford, and J.~G. Werthen, ``{A 19\% efficient AlGaAs solar
  cell with graded band gap},'' \emph{Applied Physics Letters}, vol.~47, pp.
  1319--1321, 1985. [Online]. Available:
  \url{http://scitation.aip.org/content/aip/journal/apl/47/12/10.1063/1.96266}
\BIBentrySTDinterwordspacing

\bibitem{Amano:1987es}
\BIBentryALTinterwordspacing
C.~Amano, H.~Sugiura, K.~Ando, M.~Yamaguchi, and A.~Saletes,
  ``\BIBforeignlanguage{English}{{High-efficiency Al0.3Ga0.7As solar cells
  grown by molecular beam epitaxy}},''
  \emph{\BIBforeignlanguage{English}{Applied Physics Letters}}, vol.~51,
  no.~14, pp. 1075--1077, 1987. [Online]. Available:
  \url{http://scitation.aip.org/content/aip/journal/apl/51/14/10.1063/1.98744}
\BIBentrySTDinterwordspacing

\bibitem{Soga:1995bv}
\BIBentryALTinterwordspacing
T.~Soga, T.~Kato, M.~Yang, and M.~Umeno, ``{High efficiency AlGaAs/Si
  monolithic tandem solar cell grown by metalorganic chemical vapor
  deposition},'' \emph{Journal of Applied Physics}, vol.~78, no.~6, pp.
  4196--4199, 1995. [Online]. Available:
  \url{http://scitation.aip.org/content/aip/journal/jap/78/6/10.1063/1.359880}
\BIBentrySTDinterwordspacing

\bibitem{Dimroth:2014jn}
\BIBentryALTinterwordspacing
F.~Dimroth, T.~Roesener, S.~Essig, C.~Weuffen, A.~Wekkeli, E.~Oliva, G.~Siefer,
  K.~Volz, T.~Hannappel, D.~Haussler, W.~Jager, and A.~W. Bett, ``{Comparison
  of Direct Growth and Wafer Bonding for the Fabrication of GaInP/GaAs
  Dual-Junction Solar Cells on Silicon},'' \emph{IEEE Journal of
  Photovoltaics}, vol.~4, no.~2, pp. 620--625, Feb. 2014. [Online]. Available:
  \url{http://ieeexplore.ieee.org/lpdocs/epic03/wrapper.htm?arnumber=6736084}
\BIBentrySTDinterwordspacing

\bibitem{Essig:2015dw}
\BIBentryALTinterwordspacing
S.~Essig, J.~Benick, M.~Schachtner, A.~Wekkeli, M.~Hermle, and F.~Dimroth,
  ``\BIBforeignlanguage{English}{{Wafer-Bonded GaInP/GaAs//Si Solar Cells With
  30\% Efficiency Under Concentrated Sunlight}},''
  \emph{\BIBforeignlanguage{English}{IEEE Journal of Photovoltaics}}, vol.~5,
  no.~3, pp. 977--981, Apr. 2015. [Online]. Available:
  \url{http://ieeexplore.ieee.org/lpdocs/epic03/wrapper.htm?arnumber=7045451}
\BIBentrySTDinterwordspacing

\bibitem{Hirst:2010ch}
\BIBentryALTinterwordspacing
L.~C. Hirst and N.~J. Ekins-Daukes, ``\BIBforeignlanguage{English}{{Fundamental
  losses in solar cells}},'' \emph{\BIBforeignlanguage{English}{Progress in
  Photovoltaics: Research and Applications}}, vol.~19, no.~3, pp. 286--293,
  Aug. 2010. [Online]. Available: \url{http://doi.wiley.com/10.1002/pip.1024}
\BIBentrySTDinterwordspacing

\bibitem{Yang:2014kz}
\BIBentryALTinterwordspacing
J.~Yang, Z.~Peng, D.~Cheong, and R.~Kleiman,
  ``\BIBforeignlanguage{English}{{Fabrication of High-Efficiency III-V on
  Silicon Multijunction Solar Cells by Direct Metal Interconnect}},''
  \emph{\BIBforeignlanguage{English}{IEEE Journal of Photovoltaics}}, vol.~4,
  no.~4, pp. 1149--1155, Jun. 2014. [Online]. Available:
  \url{http://ieeexplore.ieee.org/lpdocs/epic03/wrapper.htm?arnumber=6784010}
\BIBentrySTDinterwordspacing

\bibitem{Miller:2012fd}
\BIBentryALTinterwordspacing
O.~D. Miller, E.~Yablonovitch, and S.~R. Kurtz, ``{Strong Internal and External
  Luminescence as Solar Cells Approach the Shockley-Queisser Limit},''
  \emph{IEEE Journal of Photovoltaics}, vol.~2, no.~3, pp. 303--311, Jun. 2012.
  [Online]. Available:
  \url{http://ieeexplore.ieee.org/lpdocs/epic03/wrapper.htm?arnumber=6213058}
\BIBentrySTDinterwordspacing

\bibitem{Geisz:2014ht}
\BIBentryALTinterwordspacing
J.~F. Geisz, M.~A. Steiner, I.~Garcia, R.~M. France, D.~J. Friedman, and S.~R.
  Kurtz, ``{Implications of Redesigned, High-Radiative-Efficiency GaInP
  Junctions on III-V Multijunction Concentrator Solar Cells},'' \emph{IEEE
  Journal of Photovoltaics}, vol.~5, no.~1, pp. 418--424, Dec. 2014. [Online].
  Available:
  \url{http://ieeexplore.ieee.org/lpdocs/epic03/wrapper.htm?arnumber=6930722}
\BIBentrySTDinterwordspacing

\end{thebibliography}

\end{document}